\documentclass{article}

\usepackage{times}
\usepackage{graphicx} % more modern
\usepackage{subfigure}

\usepackage{natbib}

\usepackage{amsthm,amsmath,amssymb}

\usepackage{hyperref}

\usepackage[ruled]{algorithm2e}

\usepackage[accepted]{icml2014}

\icmltitlerunning{Stochastic Gradient Hamiltonian Monte Carlo}

\newcommand{\commentout}[1]{}

\newcommand{\pU}{U}
\newcommand{\psU}{\tilde{U}}

\newcommand{\sD}{\mathcal{D}}
\newcommand{\mM}{M}
\newcommand{\mB}{B}
\newcommand{\mC}{C}
\newcommand{\mI}{V}
\newcommand{\mS}{S}
\newcommand{\rR}{\mathbb{R}}
\newcommand{\temp}{\tau}
\newcommand{\opL}{\mathcal{L}}
\newcommand{\opS}{\mathcal{S}}
\newcommand{\vp}{r}
\newcommand{\trs}{T}
\newcommand{\hH}{H}
\newcommand{\dif}{d}
\newcommand{\dt}{\dif t}
\newcommand{\eps}{\epsilon}
\newcommand{\Ent}{h}

\newtheorem{thm:def}{Definition}[section]
\newtheorem{thm:thm}{Theorem}[section]
\newtheorem{thm:lemma}{Lemma}[section]
\newtheorem{thm:rmk}{Remark}[section]
\newtheorem{thm:corollary}{Corollary}[section]
\newtheorem{thm:prop}{Proposition}[section]

\begin{document}

\twocolumn[
\icmltitle{Stochastic Gradient Hamiltonian Monte Carlo}
\icmlauthor{Tianqi Chen}{tqchen@cs.washington.edu}
\icmlauthor{Emily B. Fox}{ebfox@stat.washington.edu}
\icmlauthor{Carlos Guestrin}{guestrin@cs.washington.edu}
\icmladdress{MODE Lab,  University of Washington,  Seattle, WA.}
\icmlkeywords{MCMC, Hamiltonian Monte carlo}
\vskip 0.3in
]

\begin{abstract}
Hamiltonian Monte Carlo~(HMC) sampling methods provide a mechanism for defining distant proposals with high acceptance probabilities in a Metropolis-Hastings framework, enabling more efficient exploration of the state space than standard random-walk proposals.  The popularity of such methods has grown significantly in recent years.  However, a limitation of HMC methods is the required gradient computation for simulation of the Hamiltonian dynamical system---such computation is infeasible in problems involving a large sample size or streaming data. Instead, we must rely on a noisy gradient estimate computed from a subset of the data.  In this paper, we explore the properties of such a stochastic gradient HMC approach. Surprisingly, the natural implementation of the stochastic approximation can be arbitrarily bad.  To address this problem we introduce a variant that uses second-order Langevin dynamics with a friction term that counteracts the effects of the noisy gradient, maintaining the desired target distribution as the invariant distribution.
Results on simulated data validate our theory.  We also provide an application of our methods to a classification task using neural networks and to online Bayesian matrix factorization.
\end{abstract}

\section{Introduction}
Hamiltonian Monte Carlo~(HMC)~\cite{Duane:1987HMC,NealHMC} sampling methods provide a powerful Markov chain Monte Carlo~(MCMC) sampling algorithm.  The methods define a Hamiltonian function in terms of the target distribution from which we desire samples---the \emph{potential energy}---and a \emph{kinetic energy} term parameterized by a set of ``momentum'' auxiliary variables.  Based on simple updates to the momentum variables, one simulates from a Hamiltonian dynamical system that enables proposals of distant states.  The target distribution is invariant under these dynamics; in practice, a discretization of the continuous-time system is needed necessitating a Metropolis-Hastings~(MH) correction, % \tbf{or non-MH mechanism~\cite{Sohl:2014ndbhmc}},
though still with high acceptance probability.  Based on the attractive properties of HMC in terms of rapid exploration of the state space, HMC methods have grown in popularity recently~\cite{NealHMC,Matthew:NoUTurn,Wang:2013:AdaptiveHMC}.

A limitation of HMC, however, is the necessity to compute the gradient of the potential energy function in order to simulate the Hamiltonian dynamical system.  We are increasingly faced with datasets having millions to billions of observations or where data come in as a stream and we need to make inferences online, such as in online advertising or recommender systems.  In these ever-more-common scenarios of massive batch or streaming data, such gradient computations are infeasible since they utilize the entire dataset, and thus are not applicable to ``big data'' problems.  Recently, in a variety of machine learning algorithms, we have witnessed the many successes of utilizing a noisy estimate of the gradient based on a \emph{minibatch} of data to scale the algorithms~\cite{SGD,SVI,Welling:2011SGLD}.  A majority of these developments have been in optimization-based algorithms~\cite{SGD,Nemirovski:2009}, and a question is whether similar efficiencies can be garnered by sampling-based algorithms that maintain many desirable theoretical properties for Bayesian inference.  One attempt at applying such methods in a sampling context is the recently proposed stochastic gradient Langevin dynamics (SGLD)~\cite{Welling:2011SGLD,Ahn:2012:SGFS,NIPS2013_4883}.  This method builds on first-order Langevin dynamics that do not include the crucial momentum term of HMC.

In this paper, we explore the possibility of marrying the efficiencies in state space exploration of HMC with the big-data computational efficiencies of stochastic gradients.  Such an algorithm would enable a large-scale and online Bayesian sampling algorithm with the potential to rapidly explore the posterior.
As a first cut, we consider simply applying a stochastic gradient modification to HMC and assess the impact of the noisy gradient.  We prove that the noise injected in the system by the stochastic gradient no longer leads to Hamiltonian dynamics with the desired target distribution as the stationary distribution.  As such, even before discretizing the dynamical system, we need to correct for this effect.  One can correct for the injected gradient noise through an MH step, though this itself requires costly computations on the entire dataset. In practice, one might propose long simulation runs before an MH correction, but this leads to low acceptance rates due to large deviations in the Hamiltonian from the injected noise. The efficiency of this MH step could potentially be improved using the recent results of~\cite{Korattikara:2013austerity,Bardenet:2014subset}. In this paper, we instead introduce a stochastic gradient HMC method with \emph{friction} added to the momentum update. We assume the injected noise is Gaussian, appealing to the central limit theorem, and analyze the corresponding dynamics. We show that using such \emph{second-order Langevin dynamics} enables us to maintain the desired target distribution as the stationary distribution.  That is, the friction counteracts the effects of the injected noise.  For discretized systems, we consider letting the step size tend to zero so that an MH step is not needed, giving us a significant computational advantage.  Empirically, we demonstrate that we have good performance even for $\eps$ set to a small, fixed value.  The theoretical computation versus accuracy tradeoff of this small-$\eps$ approach is provided in the Supplementary Material.

A number of simulated experiments validate our theoretical results and demonstrate the differences between (i) exact HMC, (ii) the na\"{\i}ve implementation of stochastic gradient HMC (simply replacing the gradient with a stochastic gradient), and (iii) our proposed method incorporating friction.  We also compare to the first-order Langevin dynamics of SGLD.  Finally, we apply our proposed methods to a classification task using Bayesian neural networks and to online Bayesian matrix factorization of a standard movie dataset.  Our experimental results demonstrate the effectiveness of the proposed algorithm.

%The remainder of the paper is organized as follows. Sec.~\ref{sec:hmc} introduces the HMC algorithm. Sec.~\ref{sec:sghmc} provides an analysis of cases when HMC is combined with stochastic gradient and proposes our stochastic gradient Hamiltonian Monte Carlo (SGHMC) algorithm.  Experimental results are reported in Sec.~\ref{sec:exp}.
%
\section{Hamiltonian Monte Carlo}\label{sec:hmc}
Suppose we want to sample from the posterior distribution of $\theta$ given a set of independent observations $x \in \sD$:
\begin{equation}
    p(\theta|\mathcal{D}) \propto \exp( - \pU(\theta) ),
\end{equation}
where the \emph{potential energy} function $\pU$ is given by
\begin{equation}\label{eq:exactU}
   \pU =  - \sum_{x\in\sD} \log p(x|\theta) - \log p(\theta).
\end{equation}

Hamiltonian~(Hybrid) Monte Carlo (HMC)~\cite{Duane:1987HMC,NealHMC} provides a method for proposing samples of $\theta$ in a Metropolis-Hastings (MH) framework that efficiently explores the state space as compared to standard random-walk proposals.  These proposals are generated from a Hamiltonian system based on introducing a set of auxiliary momentum variables, $\vp$.  That is, to sample from $p(\theta|\sD)$, HMC considers generating samples from a joint distribution of $(\theta,\vp)$ defined by
\begin{equation}
    \pi(\theta,\vp)  \propto \exp\left( -\pU(\theta)- \frac{1}{2}\vp^\trs \mM^{-1} \vp \right).
\label{eq:HMC-pi}
\end{equation}
If we simply discard the resulting $r$ samples, the $\theta$ samples have marginal distribution $p(\theta|\sD)$. Here, $\mM$ is a mass matrix, and together with $\vp$, defines a \emph{kinetic energy} term. $\mM$ is often set to the identity matrix, $I$, but can be used to precondition the sampler when we have more information about the target distribution. The Hamiltonian function is defined by $\hH(\theta,\vp) = \pU(\theta) + \frac{1}{2}\vp^\trs \mM^{-1} \vp$. Intuitively, $\hH$ measures the total energy of a physical system with position variables $\theta$ and momentum variables $\vp$.

To propose samples, HMC simulates the Hamiltonian dynamics
\begin{equation}\label{eq:hmc}
\left\{
\begin{array}{ll}
    \dif \theta =&  \mM^{-1} \vp \ \dif t  \\
    \dif \vp  =& - \nabla \pU(\theta)\  \dif t. \\
\end{array}
\right.
\end{equation}
To make Eq.~\eqref{eq:hmc} concrete, a common analogy in 2D is as follows~\cite{NealHMC}.  Imagine a hockey puck sliding over a frictionless ice surface of varying height.  The potential energy term is based on the height of the surface at the current puck position, $\theta$, while the kinetic energy is based on the momentum of the puck, $r$, and its mass, $M$.  If the surface is flat ($\nabla \pU(\theta) = 0, \forall \theta$), the puck moves at a constant velocity. For positive slopes ($\nabla \pU(\theta) > 0$), the kinetic energy decreases as the potential energy increases until the kinetic energy is 0 ($\vp = 0$).  The puck then slides back down the hill increasing its kinetic energy and decreasing potential energy.  Recall that in HMC, the position variables are those of direct interest whereas the momentum variables are artificial constructs (auxiliary variables).

\begin{algorithm}[t]
    \caption{Hamiltonian Monte Carlo}\label{alg:hmc}
    \KwIn{Starting position $\theta^{(1)}$ and step size $\eps$}
    \For{$t=1,2 \cdots $}{
        \emph{Resample momentum $\vp$}\\
        $\vp^{(t)} \sim \mathcal{N}( 0, \mM )$\\
        $(\theta_0, \vp_0) = (\theta^{(t)}, \vp^{(t)})$\\
        \emph{Simulate discretization of Hamiltonian dynamics in Eq.~\eqref{eq:hmc}:}\\
        $\vp_0\leftarrow \vp_0 - \frac{\eps}{2}\nabla \pU(\theta_0)$\\
        \For{$i=1$ {\bfseries to} $m$}{
            $\theta_i \leftarrow \theta_{i-1} + \eps \mM^{-1} \vp_{i-1}$\\
            $\vp_i \leftarrow \vp_{i-1}-\eps \nabla \pU(\theta_i)$
        }
        $\vp_m\leftarrow \vp_m - \frac{\eps}{2}\nabla \pU(\theta_m)$\\
        $(\hat{\theta},\hat{\vp}) = (\theta_m, \vp_m)$\\
        \emph{Metropolis-Hastings correction}:\\
		$u \sim \mbox{Uniform}[0,1]$\\
        $\rho = e^{ H(\hat{\theta},\hat{\vp}) - H( \theta^{(t)},\vp^{(t)})}$\\
        \textbf{if} $u < \min(1,\rho)$, \textbf{then} $\theta^{(t+1)} = \hat{\theta}$ %\textbf{else}  $\theta^{(t+1)} = \theta^{(t)}$.
    }
\end{algorithm}

Over any interval $s$, the Hamiltonian dynamics of Eq.~\eqref{eq:hmc} defines a mapping from the state at time $t$ to the state at time $t+s$.  Importantly, this mapping is reversible, which is important in showing that the dynamics leave $\pi$ invariant. Likewise, the dynamics preserve the total energy, $\hH$,
so proposals are always accepted. In practice, however, we usually cannot simulate exactly from the continuous system of Eq.~\eqref{eq:hmc} and instead consider a discretized system.  One common approach is the ``leapfrog'' method, which is outlined in Alg.~\ref{alg:hmc}.  Because of inaccuracies introduced through the discretization, an MH step must be implemented (i.e., the acceptance rate is no longer 1).  However, acceptance rates still tend to be high even for proposals that can be quite far from their last state.

There have been many recent developments of HMC to make the algorithm more flexible and applicable in a variety of settings. The ``No U-Turn'' sampler~\cite{Matthew:NoUTurn} and the methods proposed by~\citet{Wang:2013:AdaptiveHMC} allow automatic tuning of the step size, $\epsilon$, and number of simulation steps, $m$. Riemann manifold HMC~\cite{Girolami:LMC} makes use of the Riemann geometry to adapt the mass $\mM$, enabling the algorithm to make use of curvature information to perform more efficient sampling.  We attempt to improve HMC in an orthogonal direction focused on computational complexity, but these adaptive HMC techniques could potentially be combined with our proposed methods to see further benefits.

\section{Stochastic Gradient HMC}\label{sec:sghmc}
In this section, we study the implications of implementing HMC using a stochastic gradient and propose variants on the Hamiltonian dynamics that are more robust to the noise introduced by the stochastic gradient estimates. In all scenarios, instead of directly computing the costly gradient $\nabla \pU(\theta)$ using Eq.~\eqref{eq:exactU}, which requires examination of the entire dataset $\sD$, we consider a noisy estimate based on a \emph{minibatch} $\tilde{\sD}$ sampled uniformly at random from $\sD$:
\begin{equation}
    \nabla \psU(\theta) = -\frac{|\sD|}{|\tilde{\sD}|} \sum_{x\in\tilde{\sD}}\nabla \log p(x|\theta) - \nabla \log p(\theta),\ \tilde{\sD} \subset \sD.
\end{equation}
We assume that our observations $x$ are \emph{independent} and, appealing to the central limit theorem, approximate this noisy gradient as
\begin{equation}
    \nabla \psU(\theta) \approx \nabla \pU(\theta ) + \mathcal{N}(0, \mI(\theta)).
\label{eq:noisygrad}
\end{equation}
Here, $\mI$ is the covariance of the stochastic gradient noise, which can depend on the current model parameters and sample size. Note that we use an abuse of notation in Eq.~\eqref{eq:noisygrad} where the addition of $\mathcal{N}(\mu,\Sigma)$ denotes the introduction of a random variable that is distributed according to this multivariate Gaussian.  As the size of $\tilde{\sD}$ increases, this Gaussian approximation becomes more accurate. Clearly, we want minibatches to be small to have our sought-after computational gains.  Empirically, in a wide range of settings, simply considering a minibatch size on the order of hundreds of data points is sufficient for the central limit theorem approximation to be accurate~\cite{Ahn:2012:SGFS}. In our applications of interest, minibatches of this size still represent a significant reduction in the computational cost of the gradient.

\subsection{Na\"{\i}ve Stochastic Gradient HMC}\label{sec:hmc-noise}
The most straightforward approach to stochastic gradient HMC is simply to replace $\nabla \pU(\theta)$ in Alg.~\ref{alg:hmc} by $\nabla \psU(\theta)$. Referring to Eq.~\eqref{eq:noisygrad}, this introduces noise in the momentum update, which becomes $\Delta \vp  = - \eps \nabla \tilde{\pU}(\theta) = - \eps \nabla \pU(\theta)+ \mathcal{N}( 0, \eps^2 \mI  )$. %\tbf{Here the $\eps^2$ term in variance is given by $\eps$-scaling of the noise in Eq.\eqref{eq:noisygrad}}.
The resulting discrete time system can be viewed as an $\eps$-discretization of the following continuous stochastic differential equation:
\begin{equation}\label{eq:hmc-noise}
\left\{
\begin{array}{ll}
    \dif \theta =&  \mM^{-1} \vp \ \dif t  \\
    \dif \vp  =& - \nabla \pU(\theta)\  \dif t  + \mathcal{N}( 0, 2 \mB(\theta) \dif t ).
\end{array}
\right.
\end{equation}
Here, $\mB(\theta) = \frac{1}{2}\eps \mI(\theta)$ is the diffusion matrix contributed by gradient noise. As with the original HMC formulation, it is useful to return to a continuous time system in order to derive properties of the approach.  To gain some intuition about this setting, consider the same hockey puck analogy of Sec.~\ref{sec:hmc}.  Here, we can imagine the puck on the same ice surface, but with some random wind blowing as well. This wind may blow the puck further away than expected. Formally, as given by Corollary~\ref{cor:noisy-pi} of Theorem~\ref{thm:noisy-pi}, when $\mB$ is nonzero, $\pi(\theta,\vp)$ of Eq.~\eqref{eq:HMC-pi} is no longer invariant under the dynamics described by Eq.~\eqref{eq:hmc-noise}.
\begin{thm:thm}\label{thm:noisy-pi}
     Let $p_t(\theta, \vp)$ be the distribution of $(\theta,\vp)$ at time $t$ with dynamics governed by Eq.~\eqref{eq:hmc-noise}.
	 Define the entropy of $p_t$ as $\Ent(p_t) = - \int_{\theta,\vp} f( p_t( \theta, \vp) ) \dif \theta \dif \vp$, where $f(x) = x \ln x$. Assume $p_t$ is a distribution with density and gradient vanishing at infinity. %,\tbf{ and the gradient vanishes faster than $ \frac{1}{\ln p_t}$}.  $p_t(\theta, \vp)\rightarrow 0$ and $\nabla p_t(\theta, \vp)\rightarrow 0$ as $\|\theta\|\rightarrow \infty$ or $\|\vp\|\rightarrow \infty$.
	Furthermore, assume the gradient vanishes faster than $\frac{1}{\ln p_t}$.
	Then, the entropy of $p_t$ increases over time with rate
	\begin{multline}\label{eq:ent-rate}
	  \partial_t \Ent( p_t (\theta, \vp ) ) = \\
		\int_{\theta,\vp}f^{''}(p_t) (\nabla_{\vp} p_t(\theta,\vp))^\trs B( \theta ) \nabla_{\vp} p_t(\theta,\vp)  \dif\theta \dif \vp.
	\end{multline}
	Eq.~\eqref{eq:ent-rate} implies that $\partial_t \Ent(p_t(\theta,\vp)) \geq 0$ since $B(\theta)$ is a positive semi-definite matrix.
\end{thm:thm}
Intuitively, Theorem~\ref{thm:noisy-pi} is true because the noise-free Hamiltonian dynamics preserve entropy, while the additional noise term strictly increases entropy if we assume (i) $B(\theta)$ is positive definite (a reasonable assumption due to the normal full rank property of Fisher information) and (ii) $\nabla_{\vp} p_t(\theta,\vp) \neq 0$ for all $t$.  Then, jointly, the entropy strictly increases over time.  This hints at the fact that the distribution $p_t$ tends toward a uniform distribution, which can be very far from the target distribution $\pi$. %, as formalized in Corollary~\ref{cor:noisy-pi}, we formalize that $\pi$ is indeed not the invariant distribution.
\begin{thm:corollary}\label{cor:noisy-pi}
     The distribution $\pi(\theta,\vp)  \propto \exp\left( -\hH(\theta,\vp) \right)$ is no longer invariant under the dynamics in Eq.~\eqref{eq:hmc-noise}.
\end{thm:corollary}
%\begin{proof}
%Assume $\pi(\theta,\vp) = \exp\left( -\hH(\theta, \vp)\right ) / Z$ is invariant under Eq.~\eqref{eq:hmc-noise}. It is straightforward to verify that $\pi(\theta,\vp)$ and $\nabla \pi(\theta, \vp) = \frac{1}{Z}\exp\left( -\hH(\theta,\vp) \right)\nabla \hH(\theta,\vp) $ vanish at infinity. We also have $\nabla_{\vp} \pi(\theta,\vp) = \frac{1}{Z}\exp\left( -\hH(\theta,\vp) \right)\mM^{-1} \vp  $. Using Theorem~\ref{thm:noisy-pi}, we conclude that entropy increases over time, $\partial_t \Ent( p_t (\theta, \vp ) )|_{p_t=\pi} > 0$, which contradicts that $\pi$ is the invariant distribution.
%\end{proof}
The proofs of Theorem~\ref{thm:noisy-pi} and Corollary~\ref{cor:noisy-pi} are in the Supplementary Material.

Because $\pi$ is no longer invariant under the dynamics of Eq.~\eqref{eq:hmc-noise}, we must introduce a correction step even before considering errors introduced by the discretization of the dynamical system.  For the correctness of an MH step (based on the entire dataset), we appeal to the same arguments made for the HMC data-splitting technique of~\citet{NealHMC}.  This approach likewise considers minibatches of data and simulating the (continuous) Hamiltonian dynamics on each batch sequentially.
Importantly, \citet{NealHMC} alludes to the fact that the resulting $H$ from the split-data scenario may be far from that of the full-data scenario after simulation, which leads to lower acceptance rates and thereby reduces the apparent computational gains in simulation. Empirically, as we demonstrate in Fig.~\ref{fig:hmc-sim-trace}, we see that even finite-length simulations from the noisy system can diverge quite substantially from those of the noise-free system.  Although the minibatch-based HMC technique considered herein is slightly different from that of \citet{NealHMC}, the theory we have developed in Theorem \ref{thm:noisy-pi} surrounding the high-entropy properties of the resulting invariant distribution of Eq.~\eqref{eq:hmc-noise} provides some intuition for the observed deviations in $H$ both in our experiments and those of~\citet{NealHMC}.

The poorly behaved properties of the trajectory of $H$ based on simulations using noisy gradients results in a complex computation versus efficiency tradeoff.  On one hand, it is extremely computationally intensive in large datasets to insert an MH step after just short simulation runs (where deviations in $H$ are less pronounced and acceptance rates should be reasonable).  Each of these MH steps requires a costly computation using \emph{all} of the data, thus defeating the computational gains of considering noisy gradients.  On the other hand, long simulation runs between MH steps can lead to very low acceptance rates.  Each rejection corresponds to a wasted (noisy) gradient computation and simulation using the proposed variant of Alg.~\ref{alg:hmc}.  One possible direction of future research is to consider using the recent results of~\citet{Korattikara:2013austerity} and \citet{Bardenet:2014subset} that show that it is possible to do MH using a subset of data.  However, we instead consider in Sec.~\ref{sec:SGHMC-friction} a straightforward modification to the Hamiltonian dynamics that alleviates the issues of the noise introduced by stochastic gradients. In particular, our modification allows us to again achieve the desired $\pi$ as the invariant distribution of the continuous Hamiltonian dynamical system.

\subsection{Stochastic Gradient HMC with Friction} \label{sec:SGHMC-friction}
In Sec.~\ref{sec:hmc-noise}, we showed that HMC with stochastic gradients requires a frequent costly MH correction step, or alternatively, long simulation runs with low acceptance probabilities.  Ideally, instead, we would like to minimize the effect of the injected noise on the dynamics themselves to alleviate these problems.  To this end, we consider a modification to Eq.~\eqref{eq:hmc-noise} that adds a ``friction'' term to the momentum update:
\begin{equation}\label{eq:sghmc}
\left\{
\begin{array}{ll}
    \dif \theta \hspace{-8pt} &=  \mM^{-1} \vp \ \dt  \\
    \dif \vp\hspace{-8pt} &= - \nabla \pU(\theta)\  \dt - \mB \mM^{-1} \vp \dt + \mathcal{N}( 0, 2 \mB \dt ). \\
\end{array}
\right.
\end{equation}
Here and throughout the remainder of the paper, we omit the dependence of $B$ on $\theta$ for simplicity of notation.
Let us again make a hockey analogy. Imagine we are now playing street hockey instead of ice hockey, which introduces friction from the asphalt.  There is still a random wind blowing, however the friction of the surface prevents the puck from running far away.  That is, the friction term $\mB \mM^{-1} \vp$ helps decrease the energy $\hH(\theta,\vp)$, thus reducing the influence of the noise. This type of dynamical system is commonly referred to as \emph{second-order Langevin dynamics} in physics~\cite{Wang:1945}. Importantly, we note that the Langevin dynamics used in SGLD~\cite{Welling:2011SGLD} are first-order, which can be viewed as a limiting case of our second-order dynamics when the friction term is large.  Further details on this comparison follow at the end of this section.
\begin{thm:thm}\label{thm:shmc-stat}
   $\pi(\theta,\vp) \propto \exp(-\hH(\theta,\vp) )$ is the unique stationary distribution of the dynamics described by Eq.~\eqref{eq:sghmc}.
\end{thm:thm}
\begin{proof}
Let
$G=\left[ \begin{array}{cc}
          0  & -I\\
          I  & 0 \\
\end{array}\right]$,
$D=\left[\begin{array}{cc}
    0  & 0\\
    0  & \mB \\
\end{array}\right]$, where $G$ is an anti-symmetric matrix, and $D$ is the symmetric (diffusion) matrix.
Eq.~\eqref{eq:sghmc} can be written in the following decomposed form~\cite{Yin:JPA,Shi:JSP}
\begin{equation*}
\begin{split}
    \dif \left[ \begin{array}{c} \theta \\ \vp  \end{array} \right]
    =& -\left[  \begin{array}{cc}
            0  & -I\\
            I  & \mB \\
            \end{array}
            \right]
            \left[
            \begin{array}{c}
                \nabla \pU(\theta) \\  \mM^{-1} \vp
            \end{array} \right] \dt + \mathcal{N}(0, 2 D \dt)\\
    =& -\left[D +G\right]\nabla \hH(\theta,\vp) \dt + \mathcal{N}(0, 2  D \dt).
\end{split}
\end{equation*}
The distribution evolution under this dynamical system is governed by a Fokker-Planck equation%\footnote{This form is not standard, but is equivalent to the standard form~(proof in supplementary material). We use it to reveal the stationary distribution more clearly.}
\begin{equation}\label{eq:fpe}
\partial_t p_t(\theta,\vp)\hspace{-2pt}  =\hspace{-3pt}  \nabla^\trs \{[D + G]\left[  p_t(\theta,\vp) \nabla \hH(\theta,\vp) +  \nabla p_t(\theta,\vp) \right]\}.
\end{equation}
See the Supplementary Material for details. We can verify that $\pi(\theta,\vp)$ is invariant under Eq.~\eqref{eq:fpe} by calculating $\left[ e^{ - \hH(\theta,\vp)} \nabla \hH(\theta,\vp) + \nabla e^{ - \hH(\theta,\vp)} \right] = 0$. Furthermore, due to the existence of diffusion noise, $\pi$ is the unique stationary distribution of Eq.~\eqref{eq:fpe}.
\end{proof}
%
%Intuitively, Theorem~\ref{thm:shmc-stat} says the stationary distribution under Eq.~\eqref{eq:sghmc} is given by a Boltzmann-Gibbs distribution of position and momentum in a physical system with energy defined by $H$.
In summary, we have shown that the dynamics given by Eq.~\eqref{eq:sghmc} have a similar invariance property to that of the original Hamiltonian dynamics of Eq.~\eqref{eq:hmc}, even with noise present.  The key was to introduce a friction term using second-order Langevin dynamics.  %This idea of coping with noise via friction is well known in statistical physics.
Our revised momentum update can also be viewed as akin to partial momentum refreshment~\cite{Horowitz:1991,Neal:1992}, which also corresponds to second-order Langevin dynamics.  Such partial momentum refreshment was shown to not greatly improve HMC in the case of noise-free gradients~\cite{NealHMC}.  However, as we have demonstrated, the idea is crucial in our stochastic gradient scenario in order to counterbalance the effect of the noisy gradients.  We refer to the resulting method as \emph{stochastic gradient HMC} (SGHMC).
\subsubsection*{Connection to First-order Langevin Dynamics}
As we previously discussed, the dynamics introduced in Eq.~\eqref{eq:sghmc} relate to the first-order Langevin dynamics used in SGLD~\cite{Welling:2011SGLD}. In particular, the dynamics of SGLD can be viewed as second-order Langevin dynamics with a large friction term. To intuitively demonstrate this connection, let $\mB \mM^{-1}= \frac{1}{\dt}$ in Eq.~\eqref{eq:sghmc}. Because the friction and momentum noise terms are very large, the momentum variable $\vp$ changes much faster than $\theta$. Thus, relative to the rapidly changing momentum, $\theta$ can be considered as \emph{fixed}. We can study this case as simply:
\begin{equation}\label{eq:mom-chg}
    \dif \vp  = - \nabla \pU(\theta) \dt -  \mB\mM^{-1} \vp \dt + \mathcal{N}( 0, 2 \mB  \dt  )
\end{equation}
The fast evolution of $\vp$ leads to a rapid convergence to the stationary distribution of Eq.~\eqref{eq:mom-chg}, which is given by $ \mathcal{N}( \mM \mB^{-1} \nabla\pU(\theta), \mM ) $. Let us now consider a change in $\theta$, with $\vp \sim  \mathcal{N}( \mM \mB^{-1} \nabla\pU(\theta), \mM )$.  Recalling $\mB \mM^{-1}= \frac{1}{\dt}$, we have
\begin{equation}\label{eq:sghmc-sgld}
        \dif \theta = - \mM^{-1} \nabla \pU(\theta) \dt^2   + \mathcal{N}( 0, 2 \mM^{-1} dt^2  ),
\end{equation}
which exactly aligns with the dynamics of SGLD where $\mM^{-1}$ serves as the preconditioning matrix~\cite{Welling:2011SGLD}. Intuitively, this means that when the friction is large, the dynamics do not depend on the decaying series of past gradients represented by $\dif \vp$, reducing to first-order Langevin dynamics.

\subsection{Stochastic Gradient HMC in Practice}\label{sec:sghmc-practice}
\begin{algorithm}[t]
    \caption{Stochastic Gradient HMC}\label{alg:sghmc}
    \For{$t=1,2 \cdots $}{
        \emph{optionally, resample momentum $\vp$ as}\\
		$\vp^{(t)} \sim \mathcal{N}( 0, \mM )$\\
        $(\theta_0, \vp_0) = (\theta^{(t)}, \vp^{(t)})$\\
        \emph{simulate dynamics in Eq.\eqref{eq:sghmc-real}:}\\
        \For{$i=1$ {\bfseries to} $m$}{
            $\theta_i \leftarrow \theta_{i-1} + \eps_t \mM^{-1} \vp_{i-1}$\\
            $\hspace{-5pt}\begin{array}{ll}\vp_i \leftarrow& \hspace{-6pt} \vp_{i-1}-\eps_t \nabla \psU(\theta_i) -\eps_t \mC \mM^{-1}\vp_{i-1} \\& + \mathcal{N}( 0, 2  (\mC - \hat{\mB}) \eps_t )\end{array}$\\
        }

        $(\theta^{(t+1)},\vp^{(t+1)}) = (\theta_m, \vp_m)$, no M-H step
    }
\end{algorithm}
In everything we have considered so far, we have assumed that we know the noise model $\mB$. Clearly, in practice this is not the case.  Imagine instead that we simply have an estimate $\hat{\mB}$.  As will become clear, it is beneficial to instead introduce a user specified friction term $\mC \succeq \hat{\mB}$ and consider the following dynamics
\begin{equation}\label{eq:sghmc-real}
\left\{
\begin{array}{ll}
    \dif \theta  =& \hspace{-8pt}  \mM^{-1} \vp \ \dt  \\
    \dif \vp =& \hspace{-8pt}- \nabla \pU(\theta)\  \dt - \mC \mM^{-1} \vp \dt \\
        &\hspace{-8pt} + \mathcal{N}( 0, 2  (\mC - \hat{\mB}) \dt ) +  \mathcal{N}( 0, 2 \mB \dt )\\
\end{array}
\right.
\end{equation}
The resulting SGHMC algorithm is shown in Alg.~\ref{alg:sghmc}. Note that the algorithm is purely in terms of user-specified or computable quantities. To understand our choice of dynamics, we begin with the unrealistic scenario of perfect estimation of $\mB$.
\begin{thm:prop}
    If $\hat{\mB}=\mB$, then the dynamics of Eq.~\eqref{eq:sghmc-real} yield the stationary distribution $\pi(\theta,\vp)\propto e^{-\hH(\theta, \vp)}$.
\end{thm:prop}
\begin{proof}
    The momentum update simplifies to $\vp = -\nabla \pU(\theta)\  \dt - \mC \mM^{-1} \vp \dt + \mathcal{N}( 0, 2 \mC \dt )$, with friction term  $\mC \mM^{-1} $ and noise term $\mathcal{N}( 0, 2 \mC \dt )$. Noting that the proof of Theorem~\ref{thm:shmc-stat} only relied on a matching of noise and friction, the result follows directly by using $C$ in place of $B$ in Theorem~\ref{thm:shmc-stat}.
%Note the proof of Theorem~\ref{thm:shmc-stat} only requires the match of noise and friction term to get the correct stationary distribution. The proof is then identical to that of Theorem~\ref{thm:shmc-stat} using $C$ in place of $B$.}
\end{proof}
Now consider the benefit of introducing the $\mC$ terms and revised dynamics in the more realistic scenario of inaccurate estimation of $\mB$. For example, the simplest choice is $\hat{\mB}=0$. Though the true stochastic gradient noise $\mB$ is clearly non-zero, as the step size $\eps \rightarrow 0$, $\mB= \frac{1}{2}\eps\mI$ goes to $0$ and $\mC$ dominates. That is, the dynamics are again governed by the controllable injected noise $\mathcal{N}( 0, 2  \mC \dt )$ and friction $\mC \mM^{-1}$. It is also possible to set $\hat{\mB} = \frac{1}{2}\eps \hat{\mI}$, where $\hat{\mI}$ is estimated using empirical Fisher information as in~\cite{Ahn:2012:SGFS} for SGLD.  %\tbf{The simulation could potentially be improved using higher-order integration, which we leave as a future direction}.

\subsubsection*{Computational Complexity}
The complexity of Alg.~\ref{alg:sghmc} depends on the choice of $\mM$, $\mC$ and $\hat{\mB}$, and the complexity for estimating $\nabla \psU(\theta)$---denoted as $g(|\sD|,d)$---where $d$ is the dimension of the parameter space. Assume we allow $\hat{\mB}$ to be an arbitrary $d\times d$ positive definite matrix. Using empirical Fisher information estimation of $\hat{\mB}$, the per-iteration complexity of this estimation step is $O(d^2| \tilde{\sD} |)$. Then, the time complexity for the $(\theta, \vp)$ update is $O(d^3)$, because the update is dominated by generating Gaussian noise with a full covariance matrix. In total, the per-iteration time complexity is $O( d^2 |\tilde{\sD}|+d^3 + g(|\tilde{\sD}|, d)  )$. In practice, we restrict all of the matrices to be diagonal when $d$ is large, resulting in time complexity $O( d|\tilde{\sD}| + d  + g(|\tilde{\sD}|, d)  )$. Importantly, we note that our SGHMC time complexity is the \emph{same} as that of SGLD~\cite{Welling:2011SGLD,Ahn:2012:SGFS} in both parameter settings.

In practice, we must assume inaccurate estimation of $B$.  For a decaying series of step sizes $\eps_t$, an MH step is not required~\cite{Welling:2011SGLD,Ahn:2012:SGFS}\footnote{We note that, just as in SGLD, an MH correction is not even possible because we cannot compute the probability of the reverse dynamics.}. However, as the step size decreases, the efficiency of the sampler likewise decreases since proposals are increasingly close to their initial value.  In practice, we may want to tolerate some errors in the sampling accuracy to gain efficiency. As in~\cite{Welling:2011SGLD,Ahn:2012:SGFS} for SGLD, we consider using a small, non-zero $\eps$ leading to some bias.  We explore an analysis of the errors introduced by such finite-$\eps$ approximations in the Supplementary Material.

\subsubsection*{Connection to SGD with Momentum}
Adding a momentum term to stochastic gradient descent (SGD) is common practice. In concept, there is a clear relationship between SGD with momentum and SGHMC, and here we formalize this connection. Letting $v=\eps \mM^{-1} \vp$, we first rewrite the update rule in Alg.~\ref{alg:sghmc} as
\begin{equation}\label{eq:sghmc-normal-mid}
\left\{
\begin{array}{ll}
    \Delta \theta =& \hspace{-8pt} v \\
    \Delta v =& \hspace{-8pt}-\eps^2 M^{-1} \nabla \psU(\theta)  - \eps \mM^{-1}\mC v \\
        & + \mathcal{N}( 0, 2 \eps^3  M^{-1} (\mC - \hat{\mB})  M^{-1} ).\\
\end{array}
\right.
\end{equation}
Define $\eta = \eps^2 \mM^{-1}$, $\alpha = \eps \mM^{-1} \mC $, $\hat{\beta} = \eps \mM^{-1}\hat{\mB} $. The update rule becomes
\begin{equation}\label{eq:sghmc-normal}
\left\{
\begin{array}{ll}
    \Delta \theta =&\hspace{-8pt} v\\
    \Delta v =&\hspace{-8pt} - \eta \nabla \psU (x) - \alpha v + \mathcal{N}( 0, 2  (\alpha-\hat{\beta})\eta ).\\
\end{array}
\right.
\end{equation}
Comparing to an SGD with momentum method, it is clear from Eq.~\eqref{eq:sghmc-normal} that $\eta$ corresponds to the learning rate and $1-\alpha$ the momentum term. When the noise is removed (via $C=\hat{B}=0$), SGHMC naturally reduces to a stochastic gradient method with momentum. We can use the equivalent update rule of Eq.~\eqref{eq:sghmc-normal} to run SGHMC, and borrow experience from parameter settings of SGD with momentum to guide our choices of SGHMC settings. For example, we can set $\alpha$ to a fixed small number (e.g., $0.01$ or $0.1$), select the learning rate $\eta$, and then fix $\hat{\beta} = \eta\hat{V}/2$. A more sophisticated strategy involves using momentum scheduling~\cite{SutskeverMartensDahlHinton_icml2013}. We elaborate upon how to select these parameters in the Supplementary Material.

\section{Experiments}\label{sec:exp}
\subsection{Simulated Scenarios}
% Data
%
\begin{figure}[t]
  \centering
  \includegraphics[scale=0.48]{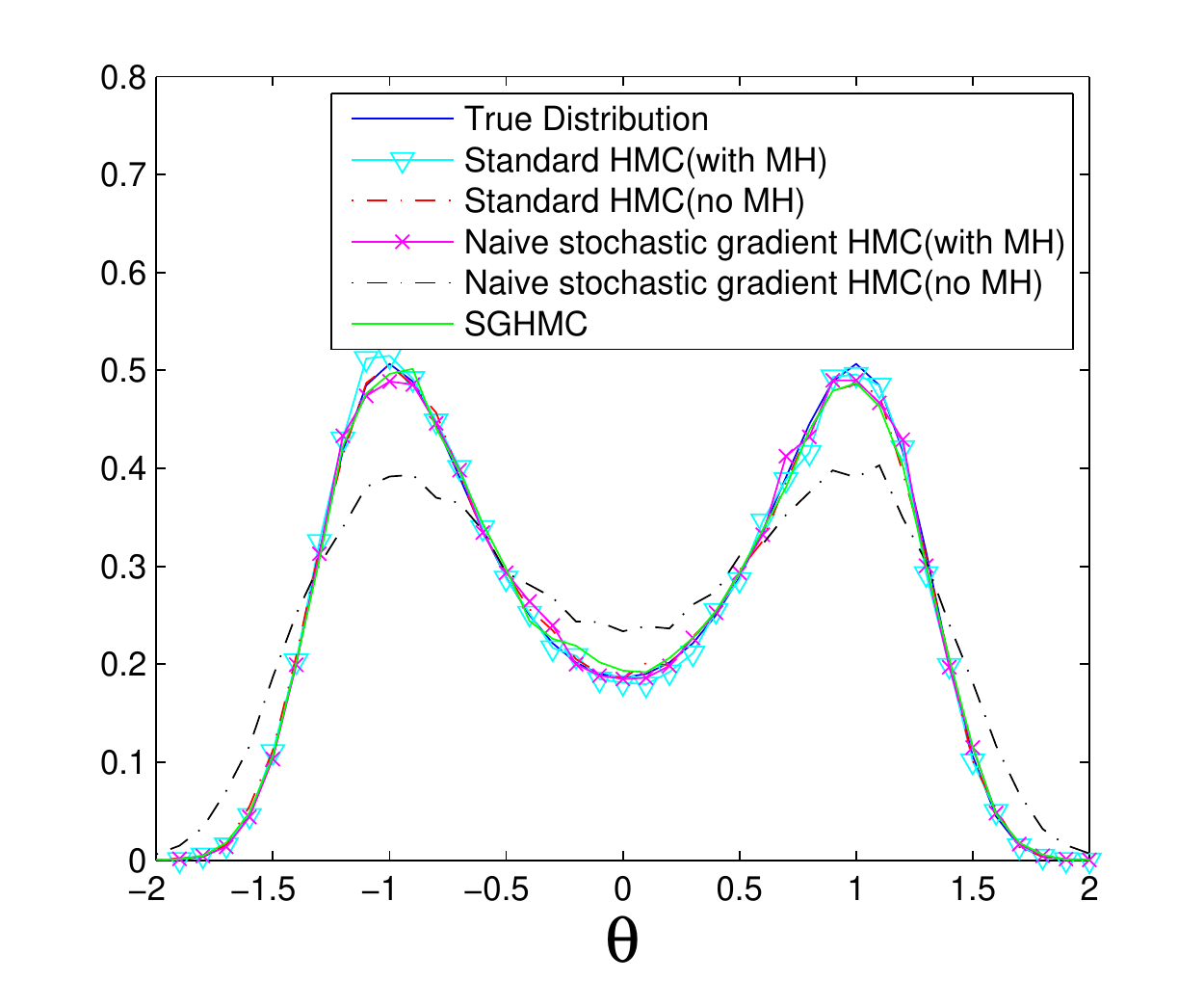}
  \vspace{-.16in}
  \caption{Empirical distributions associated with various sampling algorithms relative to the true target distribution with $U(\theta) = -2\theta^2+\theta^4$.  We compare the HMC method of Alg.~\ref{alg:hmc} with and without the MH step to: (i) a naive variant that replaces the gradient with a stochastic gradient, again with and without an MH correction; (ii) the proposed SGHMC method, which does not use an MH correction. We use $\nabla \psU(\theta) = \nabla\pU(\theta) +\mathcal{N}(0,4)$ in the stochastic gradient based samplers and $\eps=0.1$ in all cases. Momentum is resampled every 50 steps in all variants of HMC.}\label{fig:hmc-sim-samples}
\vspace{-0.16in}
\end{figure}

To empirically explore the behavior of HMC using exact gradients relative to stochastic gradients, we conduct experiments on a simulated setup. As a baseline, we consider the standard HMC implementation of Alg.~\ref{alg:hmc}, both with and without the MH correction.  We then compare to HMC with stochastic gradients, replacing $\nabla\pU$ in Alg.~\ref{alg:hmc} with $\nabla \tilde{\pU}$, and consider this proposal with and without an MH correction. Finally, we compare to our proposed SGHMC, which does not use an MH correction.  Fig.~\ref{fig:hmc-sim-samples} shows the empirical distributions generated by the different sampling algorithms. We see that even without an MH correction, both the HMC and SGHMC algorithms provide results close to the true distribution, implying that any errors from considering non-zero $\epsilon$ are negligible.  On the other hand, the results of na\"{\i}ve stochastic gradient HMC diverge significantly from the truth unless an MH correction is added.  These findings validate our theoretical results; that is, both standard HMC and SGHMC maintain $\pi$ as the invariant distribution as $\epsilon \rightarrow 0$ whereas na\"{\i}ve stochastic gradient HMC does not, though this can be corrected for using a (costly) MH step.
\begin{figure}[t]
  \centering
  \includegraphics[scale=0.45]{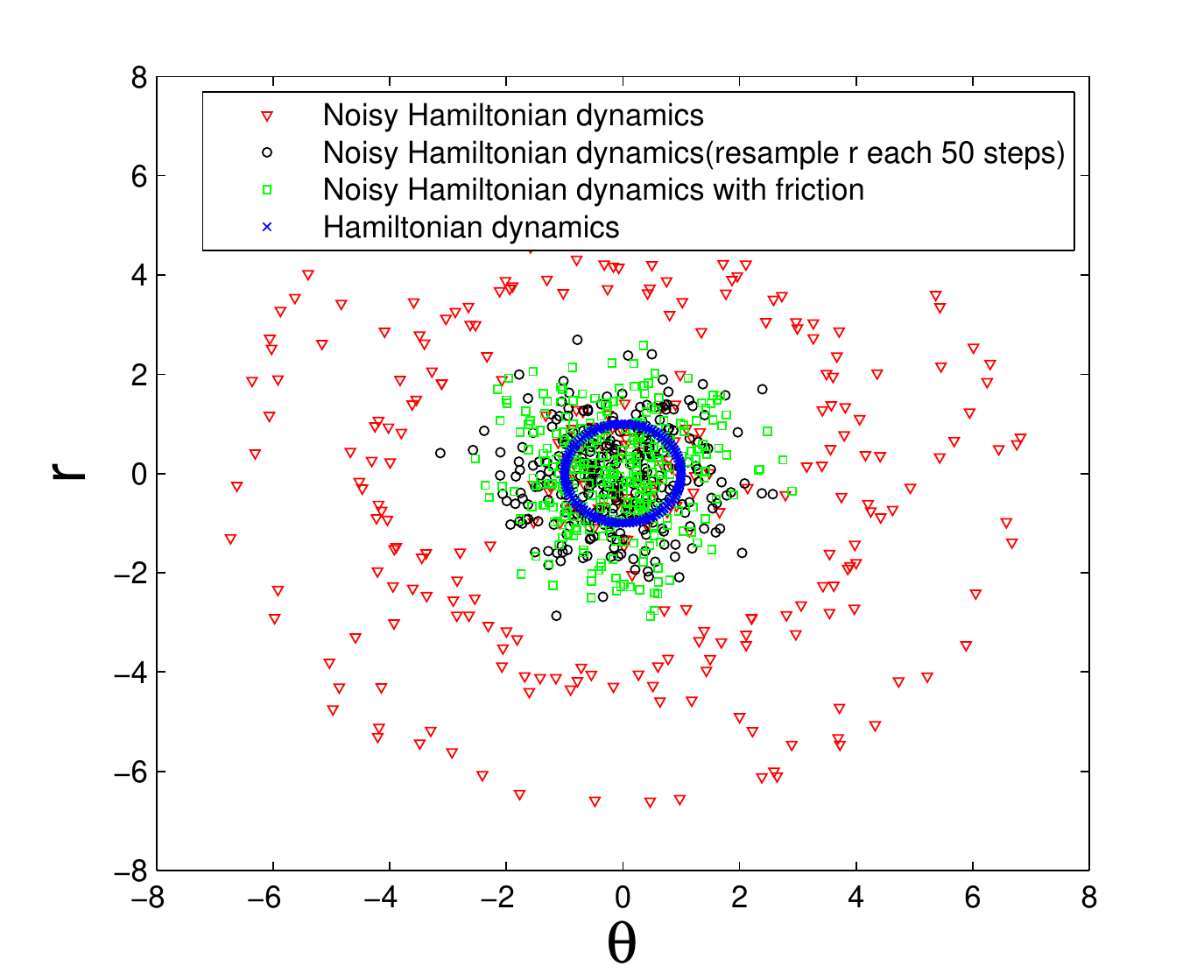}
  \vspace{-.15 in}
 \caption{Points ($\theta$,$\vp$) simulated from discretizations of various Hamiltonian dynamics over 15000 steps using $\pU(\theta) = \frac{1}{2} \theta^2$ and $\eps=0.1$. For the noisy scenarios, we replace the gradient by $\nabla \psU(\theta) = \theta +\mathcal{N}(0,4)$. We see that noisy Hamiltonian dynamics lead to diverging trajectories when friction is not introduced. Resampling $\vp$ helps control divergence, but the associated HMC stationary distribution is not correct, as illustrated in Fig.~\ref{fig:hmc-sim-samples}.}\label{fig:hmc-sim-trace}
 \vspace{-.15 in}
\end{figure}

We also consider simply simulating from the discretized Hamiltonian dynamical systems associated with the various samplers compared.  In Fig.~\ref{fig:hmc-sim-trace}, we compare the resulting trajectories and see that the path of $(\theta,\vp)$ from the noisy system \emph{without} friction diverges significantly.  The modification of the dynamical system by adding friction (corresponding to SGHMC) corrects this behavior.  We can also correct for this divergence through periodic resampling of the momentum, though as we saw in Fig.~\ref{fig:hmc-sim-samples}, the corresponding MCMC algorithm (``Naive stochastic gradient HMC (no MH)'') does not yield the correct target distribution.  These results confirm the importance of the friction term in maintaining a well-behaved Hamiltonian \emph{and} leading to the correct stationary distribution.
\begin{figure}[t]
  \centering
  \begin{tabular}{cc}
    \hspace{-0.1in}\includegraphics[scale=0.43]{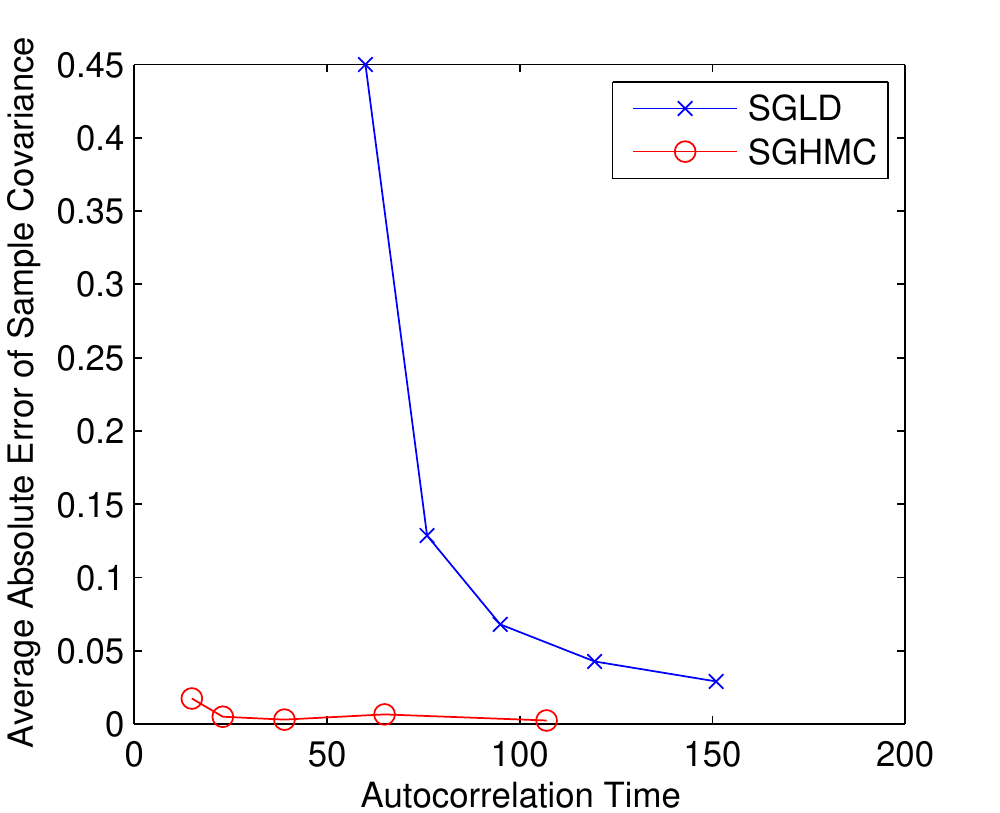} & \hspace{-0.3in}
    \includegraphics[scale=0.43]{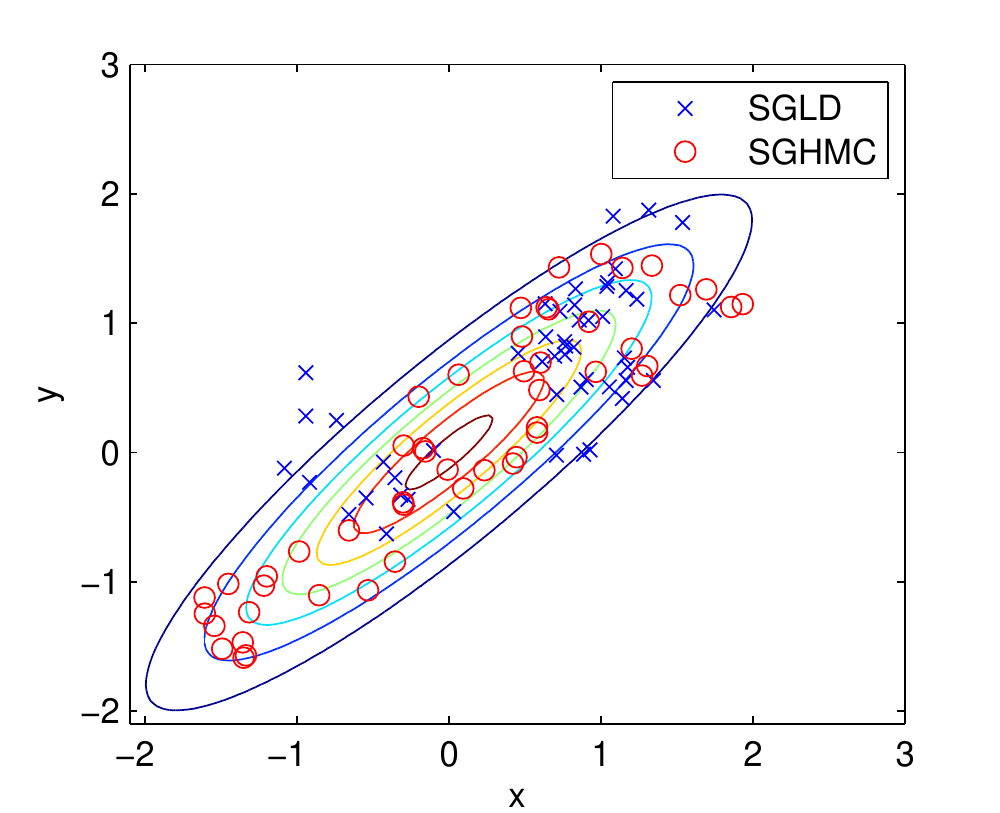}
\end{tabular}
 \vspace{-0.15 in}
 \caption{Contrasting sampling of a bivariate Gaussian with correlation using SGHMC versus SGLD. Here, $\pU(\theta) = \frac{1}{2} \theta^\trs \Sigma ^{-1} \theta$, $\nabla \psU(\theta) = \Sigma^{-1}\theta +\mathcal{N}(0,I)$ with $\Sigma_{11}=\Sigma_{22} = 1$ and correlation $\rho=\Sigma_{12}=0.9$.  \emph{Left: } Mean absolute error of the covariance estimation using ten million samples versus autocorrelation time of the samples as a function of 5 step size settings. \emph{Right: } First 50 samples of SGHMC and SGLD.} \label{fig:sgldcmp}
\vspace{-0.15in}
\end{figure}

It is known that a benefit of HMC over many other MCMC algorithms is the efficiency in sampling from correlated distributions~\cite{NealHMC}---this is where the introduction of the momentum variable shines.  SGHMC inherits this property. Fig.~\ref{fig:sgldcmp} compares SGHMC and SGLD~\cite{Welling:2011SGLD} when sampling from a bivariate Gaussian with positive correlation.  For each method, we examine five different settings of the initial step size on a linearly decreasing scale and generate ten million samples. For each of these sets of samples (one set per step-size setting), we calculate the autocorrelation time\footnote{Autocorrelation time is defined as $1+\sum_{s=1}^{\infty} \rho_s$, where $\rho_s$ is the autocorrelation at lag $s$.} of the samples and the average absolute error of the resulting sample covariance.  Fig.~\ref{fig:sgldcmp}(a) shows the autocorrelation versus estimation error for the five settings.  As we decrease the stepsize, SGLD has reasonably low estimation error but high autocorrelation time indicating an inefficient sampler.  In contrast, SGHMC achieves even lower estimation error at very low autocorrelation times, from which we conclude that the sampler is indeed efficiently exploring the distribution.  Fig.~\ref{fig:sgldcmp}(b) shows the first 50 samples generated by the two samplers. We see that SGLD's random-walk behavior makes it challenging to explore the tails of the distribution.  The momentum variable associated with SGHMC instead drives the sampler to move along the distribution contours.

\subsection{Bayesian Neural Networks for Classification}
\label{sec:BNN}
We also test our method on a handwritten digits classification task using the MNIST dataset\footnote{http://yann.lecun.com/exdb/mnist/}. The dataset consists of 60,000 training instances and 10,000 test instances. We randomly split a validation set containing 10,000 instances from the training data in order to select training parameters, and use the remaining 50,000 instances for training. For classification, we consider a two layer Bayesian neural network with 100 hidden variables using a sigmoid unit and an output layer using softmax. We tested four methods: SGD, SGD with momentum, SGLD and SGHMC.
For the optimization-based methods, we use the validation set to select the optimal regularizer $\lambda$ of network weights\footnote{We also tried MAP inference for selecting $\lambda$ in the optimization-based method, but found similar performance.}. For the sampling-based methods, we take a fully Bayesian approach and place a weakly informative gamma prior on each layer's weight regularizer $\lambda$. The sampling procedure is carried out by running SGHMC and SGLD using minibatches of 500 training instances, then resampling hyperparameters after an entire pass over the training set. We run the samplers for 800 iterations (each over the entire training dataset) and discard the initial 50 samples as burn-in.

\begin{figure}[t]
  \centering
    \includegraphics[scale=0.65]{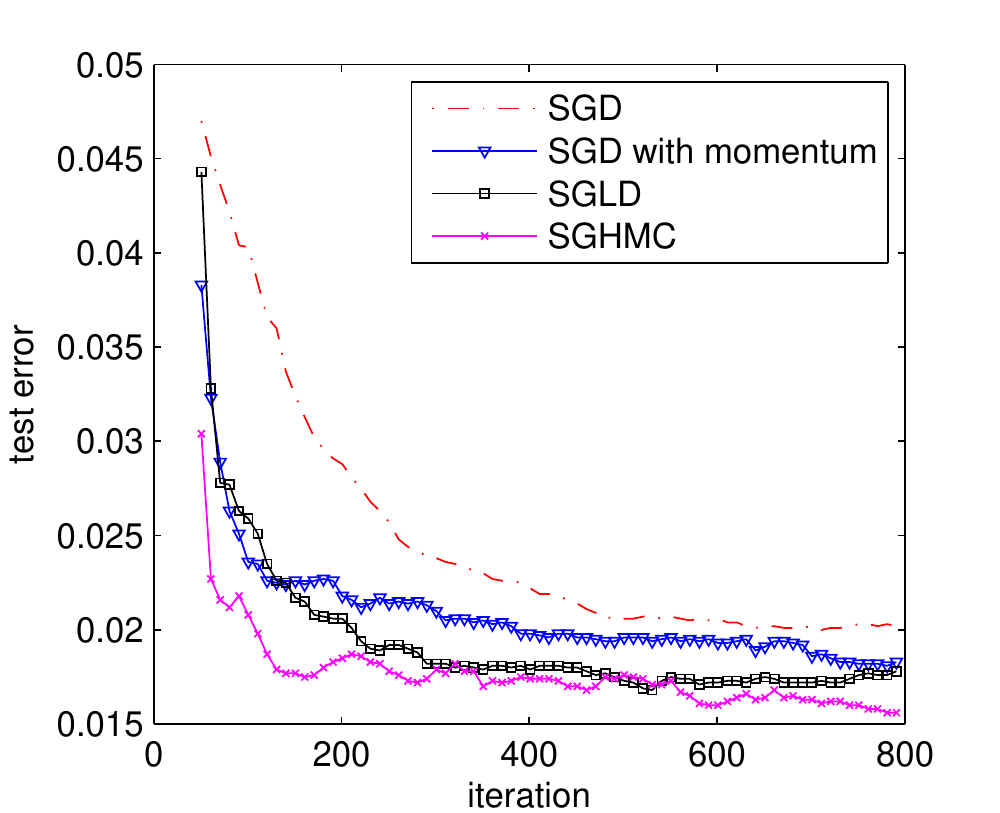}
\vspace{-0.15 in}
  \caption{Convergence of test error on the MNIST dataset using SGD, SGD with momentum, SGLD, and SGHMC to infer model parameters of a Bayesian neural net.}\label{fig:mnist}
\vspace{-0.15 in}
\end{figure}

The test error as a function of MCMC or optimization iteration (after burn-in) is reported for each of these methods in Fig.~\ref{fig:mnist}. From the results, we see that SGD with momentum converges faster than SGD. SGHMC also has an advantage over SGLD, converging to a low test error much more rapidly. In terms of runtime, in this case the gradient computation used in backpropagation dominates so both have the same computational cost. The final results of the sampling based methods are better than optimization-based methods, showing an advantage to Bayesian inference in this setting, thus validating the need for scalable and efficient Bayesian inference algorithms such as SGHMC.

\subsection{Online Bayesian Probabilistic Matrix Factorization for Movie Recommendations}
\label{sec:BPMF}
Collaborative filtering is an important problem in web applications. The task is to predict a user's preference over a set of items~(e.g., movies, music) and produce recommendations. Probabilistic matrix factorization (PMF)~\cite{SalMnih2008PMF} has proven effective for this task. Due to the sparsity in the ratings matrix (users versus items) in recommender systems, over-fitting is a severe issue with Bayesian approaches providing a natural solution~\cite{SalMnih2008BPMF}.

We conduct an experiment in \emph{online} Bayesian PMF on the Movielens dataset ml-1M\footnote{http://grouplens.org/datasets/movielens/}. The dataset contains about 1 million ratings of 3,952 movies by 6,040 users.  The number of latent dimensions is set to 20. In comparing our stochastic-gradient-based approaches, we use minibatches of 4,000 ratings to update the user and item latent matrices. We choose a significantly larger minibatch size in this application than that of the neural net because of the dramatically larger parameter space associated with the PMF model.  For the optimization-based approaches, the hyperparameters are set using cross validation (again, we did not see a performance difference from considering MAP estimation).  For the sampling-based approaches, the hyperparameters are updated using a Gibbs step after every $2,000$ steps of sampling model parameters.  We run the sampler to generate 2,000,000 samples, with the first 100,000 samples discarded as burn-in.  We use five-fold cross validation to evaluate the performance of the different methods.

\begin{table}[t]
\caption{Predictive RMSE estimated using 5-fold cross validation on the Movielens dataset for various approaches of inferring parameters of a Bayesian probabilistic matrix factorization model.}
\label{tbl:ml1m}
\vskip 0.15 in
\begin{center}
\begin{small}
\begin{sc}
\begin{tabular}{ll}
\hline
\abovespace\belowspace
Method & RMSE \\
\hline
\abovespace
SGD       & 0.8538 $\pm$ 0.0009 \\
SGD with momentum & 0.8539 $\pm$ 0.0009 \\
SGLD      & 0.8412 $\pm$ 0.0009\\
\belowspace
SGHMC     & 0.8411 $\pm$ 0.0011 \\
\hline
\end{tabular}
\end{sc}
\end{small}
\end{center}
\vskip -0.2 in
\end{table}

The results are shown in Table~\ref{tbl:ml1m}. Both SGHMC and SGLD give better prediction results than optimization-based methods.  In this experiment, the results for SGLD and SGHMC are very similar. We also observed that the per-iteration running time of both methods are comparable.  As such, the experiment suggests that SGHMC is an effective candidate for online Bayesian PMF.

\section{Conclusion}\label{sec:con}
Moving between modes of a distribution is one of the key challenges for MCMC-based inference algorithms.  To address this problem in the large-scale or online setting, we proposed SGHMC, an efficient method for generating high-quality, ``distant'' steps in such sampling methods.  Our approach builds on the fundamental framework of HMC, but using stochastic estimates of the gradient to avoid the costly full gradient computation.  Surprisingly, we discovered that the natural way to incorporate stochastic gradient estimates into HMC can lead to divergence and poor behavior both in theory and in practice.  To address this challenge, we introduced second-order Langevin dynamics with a friction term that counteracts the effects of the noisy gradient, maintaining the desired target distribution as the invariant distribution of the continuous system. Our empirical results, both in a simulated experiment and on real data, validate our theory and demonstrate the practical value of introducing this simple modification.  A natural next step is to explore combining adaptive HMC techniques with SGHMC.  More broadly, we believe that the unification of efficient optimization and sampling techniques, such as those described herein, will enable a significant scaling of Bayesian methods.

{\small
\vspace{-0.05in}
\section*{Acknowledgements}
\vspace{-0.05in}
This work was supported in part by the TerraSwarm Research Center sponsored by MARCO and DARPA, ONR Grant N00014-10-1-0746, DARPA Grant FA9550-12-1-0406 negotiated by AFOSR, NSF IIS-1258741 and Intel ISTC Big Data. We also appreciate the discussions with Mark Girolami, Nick Foti, Ping Ao and Hong Qian.
}

\bibliography{SGHMC}
\bibliographystyle{icml2014}

\newpage
\section*{Supplementary Material}
\appendix
\section{Background on Fokker-Planck Equation}
The Fokker-Planck equation~(FPE) associated with a given stochastic differential equation~(SDE) describes the time evolution of the distribution on the random variables under the specified stochastic dynamics. For example, consider the SDE:
\begin{equation}\label{eq:sde-general}
    \dif z = g( z ) dt + \mathcal{N}(0, 2 D(z) \dif t),
\end{equation}
where $z\in \rR^n, g(z) \in \rR^n, D(z) \in \rR^{n\times n}$. The distribution of $z$ governed by Eq.~\eqref{eq:sde-general}~(denoted by $p_t(z)$),  evolves under the following equation
\begin{small}
\begin{equation*}
    \partial_t p_t(z) = - \sum_{i=1}^n \partial_{z_i} [ g_i(z) p_t(z) ]  +\sum_{i=1}^n\sum_{j=1}^n\partial_{z_i}\partial_{z_j} [ D_{ij}(z) p_t(z) ].
\end{equation*}
\end{small}
\hspace{-5pt} Here $g_i(z)$ is the $i$-th entry of vector $g(z)$ and $D_{ij}(z)$ is the $(i,j)$ entry of the matrix $D$. In the dynamics considered in this paper, $z=(\theta, \vp)$ and
\begin{equation}
D=\left[\begin{array}{cc}
    0  & 0\\
    0  & \mB(\theta) \\
\end{array}\right].
\end{equation}
That is, the random variables are momentum $\vp$ and position $\theta$, with noise only added to $\vp$ (though dependent upon $\theta$). The FPE can be written in the following compact form:
\begin{equation}\label{eq:fpe-compact}
    \partial_t p_t(z) = - \nabla^\trs [ g(z) p_t(z) ]  +\nabla^\trs[ D(z) \nabla p_t(z) ],
\end{equation}
where $\nabla^\trs [ g(z) p_t(z) ] = \sum_{i=1}^n \partial_{z_i} [ g_i(z) p_t(z) ]   \ $, and
\begin{small}
\begin{equation*}
\begin{split}
&\nabla^\trs[ D \nabla p_t(\theta,\vp)]\\
&\quad= \sum_{ij} \partial_{z_{i}} [ D_{ij}( z ) \partial_{z_{j}}  p_t(z) ] \\
&\quad= \sum_{ij}\partial_{z_{i}}[ D_{ij}( z ) \partial_{z_{j}} p_t(z)] + \sum_{ij} \partial_{z_{i}}[ (  \partial_{z_{j}} D_{ij}( z ))  p_t(z)] \\
&\quad=\sum_{ij}\partial_{z_{i}}\partial_{z_{j}} [ D_{ij}( z ) p_t(z)]. \\
\end{split}
\end{equation*}
\end{small}
\hspace{-5pt} Note that $ \partial_{z_{j}} D_{ij}( z ) = 0$ for all $i,j$, since $\partial_{\vp_{j}} B_{ij}( \theta ) = 0 $ (the noise is only added to $\vp$ and only depends on parameter $\theta$).
\section{Proof of Theorem \ref{thm:noisy-pi}}
Let
$G=\left[ \begin{array}{cc}
          0  & -I\\
          I  & 0 \\
\end{array}\right]$
 and $D=\left[\begin{array}{cc}
    0  & 0\\
    0  & \mB(\theta) \\
\end{array}\right]$.
The noisy Hamiltonian dynamics of Eq.~\eqref{eq:hmc-noise} can be written as
\begin{equation*}
\begin{split}
    \dif \left[ \begin{array}{c} \theta \\ \vp  \end{array} \right]
    =& -\left[  \begin{array}{cc}
            0  & -I\\
            I  & 0 \\
            \end{array}
            \right]
            \left[
            \begin{array}{c}
                \nabla \pU(\theta) \\  \mM^{-1} \vp
            \end{array} \right] \dt + \mathcal{N}(0, 2 D \dt)\\
    =& - G\nabla \hH(\theta,\vp) \dt + \mathcal{N}(0, 2 D \dt).
\end{split}
\end{equation*}
Applying Eq.~\eqref{eq:fpe-compact}, defining $g(z)= -G\nabla \hH$), the corresponding FPE is given by
\begin{equation}\label{eq:fpe-hmc}
\partial_t p_t(\theta,\vp)\hspace{-2pt} =\hspace{-3pt} \nabla^\trs [G \nabla \hH(\theta,\vp) p_t(\theta,\vp)]  + \nabla^\trs[ D \nabla p_t(\theta,\vp)].
\end{equation}
We use $z=(\theta, \vp)$ to denote the joint variable of position and momentum. The entropy is defined by $\Ent(p_t(\theta, \vp))= - \int_{\theta,\vp} f( p_t( \theta, \vp) ) \dif \theta \dif \vp$. Here $f(x)=x \ln x$ is a strictly convex function defined on $(0,+\infty)$. The evolution of the entropy is governed by
\begin{small}
\begin{equation*}\label{eq:ent-evolution}
\begin{split}
 \partial_t \Ent( p_t(z)) =& \partial_t \int_{z} f( p_t(z) ) \dif z\\
=& - \int_{z}f'(p_t(z)) \partial_t p_t(z)\dif z \\
=& - \int_{z}f'(p_t(z)) \nabla^\trs [G \nabla \hH(z) p_t(z) ]\dif z \\& \ \ \ \ - \int_{z}f'(p) \nabla^\trs [ D(z) \nabla p_t(z)]\dif z.
\end{split}
\end{equation*}
\end{small}
\hspace{-7.5pt} The entropy evolution can be described as the sum of two parts: the noise-free Hamiltonian dynamics and the stochastic gradient noise term. The Hamiltonian dynamics part does not change the entropy, since
\begin{small}
\begin{equation*}\label{eq:ent-evolution-hmc}
\begin{split}
& - \int_{z}f'(p_t(z)) \nabla^\trs [G \nabla \hH(z) p_t ]\dif z \\
&\quad = - \int_{z}f'(p_t(z)) \nabla^\trs [G \nabla \hH(z) ] p_t \dif z \\
    &\quad \ \ \ \  -  \int_{z}f'(p_t(z)) (\nabla p_t(z))^\trs [G \nabla \hH(z)] \dif z\\
&\quad =  -\int_{z} (\nabla f(p_t(z)) )^{\trs} [G \nabla \hH(z)] \dif z\\
&\quad =  \int_{z}  f(p_t(z)) \nabla^T[G \nabla \hH(z)] \dif z = 0.
\end{split}
\end{equation*}
\end{small}
\hspace{-5pt} In the second equality, we use the fact that $\nabla^T[G \nabla \hH(z)] = - \partial_{\theta}\partial_{\vp}H +\partial_{\vp}\partial_{\theta} H = 0$. The last equality is given by integration by parts, using the assumption that the probability density vanishes at infinity and $f(x)\rightarrow 0$ as $x\rightarrow 0$ such that $ f(p_t(z)) [G \nabla \hH(z)]\rightarrow 0$ as $z\rightarrow \infty$.

The contribution due to the stochastic gradient noise can be calculated as
\begin{equation*}\label{eq:ent-evolution-diffusion}
\begin{split}
&  - \int_{z}f'(p_t(z)) \nabla^\trs[ D(z) \nabla  p_t(z)]\dif z \\
&\quad=  \int_{z} ( f^{''}(p_t(z))\nabla p_t(z) )^\trs  D(z) \nabla  p_t(z) \dif z \\
&\quad= \int_{\theta,\vp} f^{''}(p_t(z))  (\nabla_{\vp} p_t(\theta,\vp))^\trs  B(\theta) \nabla_{\vp}  p_t(\theta,\vp) \dif \theta\dif \vp.
\end{split}
\end{equation*}
The first equality is again given by integration by parts, assuming that the gradient of $p_t$ vanishes at infinity faster than $\frac{1}{\ln p_t(z)}$. That is, $f'(p_t(z))\nabla p_t(z) = (1 + \ln p_t(z) ) \nabla p_t(z)\rightarrow 0$ such that $f'(p_t(z)) [ D(z) \nabla  p_t(z)]\rightarrow 0$ as $z\rightarrow \infty$.
The statement of Theorem~\ref{thm:noisy-pi} immediately follows.

\section{Proof of Corollary~\ref{cor:noisy-pi}}
Assume $\pi(\theta,\vp) = \exp\left( -\hH(\theta, \vp)\right ) / Z$ is invariant under Eq.~\eqref{eq:hmc-noise} and is a well-behaved distribution such that \mbox{$H(\theta, \vp)\rightarrow \infty$} as $\|\theta\|,\|\vp\|\rightarrow \infty$. Then it is straightforward to verify that $\pi(\theta,\vp)$ and $\ln \pi(\theta,\vp) \nabla \pi(\theta, \vp) = \frac{1}{Z}\exp\left( -\hH(\theta,\vp) \right)\nabla \hH^2(\theta,\vp) $ vanish at infinity, such that $\pi$ satisfies the conditions of Theorem~\ref{thm:noisy-pi}. We also have $\nabla_{\vp} \pi(\theta,\vp) = \frac{1}{Z}\exp\left( -\hH(\theta,\vp) \right)\mM^{-1} \vp  $. Using the assumption that the Fisher information matrix $B(\theta)$ has full rank, and noting that $f''(p) > 0$ for $p>0$, from Eq.~\eqref{eq:ent-rate} of Theorem~\ref{thm:noisy-pi} we conclude that entropy increases over time: $\partial_t \Ent( p_t (\theta, \vp ) )|_{p_t=\pi} > 0$.  This contradicts that $\pi$ is the invariant distribution.

\section{FPE for Second-Order Langevin Dynamics}
Second-order Langevin dynamics can be described by the following equation
\begin{equation}\label{eq:2lang}
\begin{split}
    \dif \left[ \begin{array}{c} \theta \\ \vp  \end{array} \right]
    =& -\left[  \begin{array}{cc}
            0  & -I\\
            I  & \mB \\
            \end{array}
            \right]
            \left[
            \begin{array}{c}
                \nabla \pU(\theta) \\  \mM^{-1} \vp
            \end{array} \right] \dt + \mathcal{N}(0, 2 \temp D \dt)\\
    =& -\left[D +G\right]\nabla \hH(\theta,\vp) \dt + \mathcal{N}(0, 2\temp D \dt),
\end{split}
\end{equation}
where $\temp$ is a temperature (usually set to 1). In this paper, we use the following compact form of the FPE to calculate the distribution evolution under Eq~\eqref{eq:2lang}:
\begin{equation}\label{eq:fpe-ao}
\partial_t p_t(\theta,\vp)\hspace{-2pt}  =\hspace{-3pt}  \nabla^\trs \{[D + G]\left[   p_t(\theta,\vp) \nabla \hH(\theta,\vp) + \temp \nabla  p_t(\theta,\vp)\right]\}.
\end{equation}
To derive this FPE, we apply Eq.~\eqref{eq:fpe-compact} to Eq~\eqref{eq:2lang}, defining $g(z)=-(D+G)\nabla \hH$, which yields
\begin{small}
\begin{equation*}
\partial_t p_t(\theta,\vp)\hspace{-2pt}  =\hspace{-4pt}  \nabla^\trs \{[D + G]\left[ \nabla \hH(\theta,\vp) p_t(\theta,\vp) \right]\} +
 \nabla^{\trs} \left[\temp D \nabla p_t(\theta, \vp) \right].
\end{equation*}
\end{small}
\hspace{-5pt} Using the fact that $\nabla^{\trs} \left[ G \nabla p_t(\theta, \vp) \right] =- \partial_{\theta}\partial_{\vp}p_t(\theta,\vp) +\partial_{\vp}\partial_{\theta} p_t(\theta,\vp) = 0$, we get Eq.~\eqref{eq:fpe-ao}. This form of the FPE allows easy verification that the stationary distribution is given by $\pi(\theta,\vp)\propto  e^{ - \frac{1}{\temp}\hH(\theta,\vp)}$. In particular, if we substitute the target distribution into Eq.~\eqref{eq:fpe-ao}, we note that
\begin{equation*}
\begin{split}
\left[  e^{ - \frac{1}{\temp}\hH(\theta,\vp)} \nabla \hH(\theta,\vp) + \temp \nabla  e^{ - \frac{1}{\temp}}\hH(\theta,\vp) \right] = 0
\end{split}
\end{equation*}
such that $\partial_t \pi(\theta,\vp) = 0$, implying that $\pi$ is indeed the stationary distribution.

The compact form of Eq.~\eqref{eq:fpe-ao} can also be used to construct other stochastic processes with the desired invariant distribution. A generalization of the FPE in Eq.~\eqref{eq:fpe-ao} is given by \citet{Yin:JPA}. The system we have discussed in this paper considers cases where
$G=\left[ \begin{array}{cc}
            0  & -I\\
            I  & 0 \\
          \end{array} \right]$ and $D$ only depends on $\theta$.
In practice, however, it might be helpful to make $G$ depend on $\theta$ as well. For example, to make use of the Riemann geometry of the problem, as in \citet{Girolami:LMC} and \citet{NIPS2013_4883}, by adapting $G$ according to the local curvature. For us to consider these more general cases, a correction term needs to be added during simulation~\cite{Shi:JSP}.  With that correction term, we still maintain the desired target distribution as the stationary distribution.

\section{Reversibility of SGHMC Dynamics}
The dynamics of SGHMC are not reversible in the conventional definition of reversibility. However, the dynamics satisfy the following property:
\begin{thm:thm}\label{thm:shmc-db}
   Assume $ P(\theta_t,\vp_t|\theta_0,\vp_0)$ is the distribution governed by dynamics in Eq.~\eqref{eq:2lang}, i.e. $P(\theta_t,\vp_t|\theta_0,\vp_0) $ follows Eq.~\eqref{eq:fpe-ao}, then for $\pi(\theta,\vp)  \propto \exp( -\hH(\theta, \vp ) )$,
   \begin{equation}
        \pi(\theta_0,\vp_0) P(\theta_t,\vp_t|\theta_0,\vp_0) = \pi(\theta_t,-\vp_t) P(\theta_0, - \vp_0|\theta_t, -\vp_t).
   \end{equation}
\end{thm:thm}
\begin{proof}
Assuming $\pi$ is the stationary distribution and $P^*$ the reverse-time Markov process associated with $P$:  $\pi(\theta_0,\vp_0) P(\theta_t,\vp_t|\theta_0,\vp_0) = \pi(\theta_t,\vp_t) P^*(\theta_0, \vp_0|\theta_t, \vp_t)$. Let $\opL(p) =   \nabla^\trs \{[D + G]\left[ p\nabla \hH(\theta,\vp) + \temp \nabla p \right] \} $ be the generator of Markov process described by Eq.~\eqref{eq:fpe-ao}. The generator of the reverse process is given by $\opL^*$, which is the adjoint operator of $\opL$ in the inner-product space $l^2(\pi)$, with inner-product defined by  $\langle p, q \rangle_{\pi} = E_{x\sim \pi(x)}[ p(x)q(x)]$. We can verify that $\opL^*(p) = \nabla^\trs \{[D - G]\left[ p \nabla \hH(\theta,\vp) + \temp \nabla p \right] \} $. The corresponding SDE of the reverse process is given by
\begin{equation*}
d\left[ \begin{array}{c} \theta \\ \vp  \end{array} \right]
 = \left[D -G\right]\nabla \hH(\theta,\vp) + \mathcal{N}(0, 2 \temp D dt),
\end{equation*}
which is equivalent to
\begin{equation*}
d\left[ \begin{array}{c} \theta \\ - \vp  \end{array} \right]
 = \left[D +G \right]\nabla \hH(\theta,-\vp) + \mathcal{N}(0, 2 \temp D dt).
\end{equation*}
This means $ P^*(\theta_0, \vp_0|\theta_t, \vp_t) = P(\theta_0, -\vp_0|\theta_t, -\vp_t)$. Recalling that we assume Gaussian momentum, $\vp$, centered about 0, we also have $\pi(\theta,\vp) = \pi(\theta,-\vp)$.  Together, we then have
\begin{equation*}
\begin{split}
    \pi(\theta_0,\vp_0) P(\theta_t,\vp_t|\theta_0,\vp_0) &= \pi(\theta_t,\vp_t) P^*(\theta_0, \vp_0|\theta_t, \vp_t) \\
                &= \pi(\theta_t, -\vp_t) P(\theta_0, -\vp_0|\theta_t, -\vp_t).
\end{split}
\end{equation*}
\end{proof}
Theorem~\ref{thm:shmc-db} is not strictly detailed balance by the conventional definition since $\opL^*\neq \opL$ and $P^*\neq P$. However, it can be viewed as a kind of time reversibility. When we reverse time, the sign of speed needs to be reversed to allow backward travel. This property is shared by the noise-free HMC dynamics of~\cite{NealHMC}. Detailed balance can be enforced by the symmetry of $\vp$ during the re-sampling step. However, we note that we do not rely on detailed balance to have $\pi$ be the stationary distribution of our noisy Hamiltonian with friction (see Eq.~\eqref{eq:sghmc}).

\section{Convergence Analysis}\label{sec:converge}
In the paper, we have discussed that the efficiency of SGHMC decreases as the step size $\eps$ decreases. In practice, we usually want to trade a small amount of error for efficiency. In the case of SGHMC, we are interested in a small, nonzero $\eps$ and fast approximation of $\mB$ given by $\hat{\mB}$. In this case, even under the continuous dynamics, the sampling procedure contains error that relates to $\eps$ due to inaccurate estimation of $\mB$ with $\hat{\mB}$. In this section, we investigate how the choice of $\eps$ can be related to the error in the final stationary distribution. The sampling procedure with inaccurate estimation of $\mB$ can be described with the following dynamics
\begin{equation*}
\left\{
\begin{array}{ll}
    \dif \theta =&  \hspace{-9pt}  \mM^{-1} \vp \ \dt  \\
    \dif \vp =&  \hspace{-9pt} - \nabla \pU(\theta)\  \dt - \mC \mM^{-1} \vp \dt
        + \mathcal{N}( 0, 2 (\mC +\delta \mS) \dt ).
\end{array}
\right.
\end{equation*}
Here, $\delta \mS =\mB - \hat{\mB}$ is the error term that is not considered by the sampling algorithm. Assume the setting where $\hat{\mB} = 0$, then we can let $\delta=\eps$ and $\mS = \frac{1}{2} \mI$.  Let $\tilde{\pi}$ be the stationary distribution of the dynamics. In the special case when $\mI=\mC$, we can calculate $\tilde{\pi}$ exactly by
\begin{equation}
    \tilde{\pi}( \theta, \vp )  \propto \exp\left( -\frac{1}{1+\delta} \hH( \theta, \vp) \right).
\end{equation}
This indicates that for small $\epsilon$, our stationary distribution is indeed close to the true stationary distribution.  In general case, we consider the FPE of the distribution of this SDE, given by
\begin{equation}\label{eq:fpe-noise}
    \partial_t \tilde{p}_t(\theta,\vp) \hspace{-2pt}= [\opL +\delta \opS]  \tilde{p}_t(\theta,\vp).
\end{equation}
Here, $\opL(p) =   \nabla^\trs \{[D + G]\left[ p \nabla \hH(\theta,\vp) + \nabla p \right] \} $ is the operator corresponds to correct sampling process. Let the operator $\opS(p) =  \nabla_{\vp} [ \mS \nabla_{\vp} p ]$ correspond to the error term introduced by inaccurate $\hat{\mB}$. Let us consider the $\chi^2$-divergence defined by
\begin{equation*}
    \chi^2( p, \pi ) = E_{x\sim \pi}\left[  \frac{( p(x) - \pi(x) )^2}{\pi^2(x)} \right]  =  E_{x\sim \pi}\left[  \frac{p^2(x)}{\pi^2(x)} \right]-1,
\end{equation*}
which provides a measure of distance between the distribution $p$ and the true distribution $\pi$. Theorem~\ref{thm:shmc-error-bound} shows that the $\chi^2$-divergence decreases as $\delta$ becomes smaller.

\begin{thm:thm}\label{thm:shmc-error-bound}
    Assume $p_t$ evolves according to $\partial_t p_t = \opL p_t$, and satisfies the following mixing rate $\lambda$ with respect to $\chi^2$ divergence at $\tilde{\pi}$:
    $\partial_t \chi^2( p_t, \pi )|_{p_t = \tilde{\pi} } \leq - \lambda \chi^2( \tilde{\pi}, \pi )$. Further assume the process governed by $\opS$~($\partial_t q_t = \opS q_t$) has bounded divergence change $ |\partial_t \chi^2( q_t, \pi ) | < c$. Then  $\tilde{\pi}$ satisfies
    \begin{equation}
       \chi^2( \tilde{\pi}, \pi ) < \frac{\delta c}{\lambda}.
    \end{equation}
\end{thm:thm}
\begin{proof}
    Consider the divergence change of $\tilde{p}$ governed by Eq.(\ref{eq:fpe-noise}). It can be decomposed into two components, the change of divergence due to $\opL$, and the change of divergence due to $\delta \opS$
\begin{small}
    \begin{equation*}
    \begin{split}
        \partial_t \chi^2( \tilde{p}_t, \pi ) = & E_{x\sim \pi}\left[  \frac{\tilde{p}(x)}{\pi^2(x)} [\opL + \delta \opS] \tilde{p}_t(x)  \right] \\
        =&E_{x\sim \pi}\left[  \frac{\tilde{p}_t(x)}{\pi^2(x)} \opL\tilde{p}_t(x)  \right] + \delta E_{x\sim \pi}\left[  \frac{\tilde{p}(x)}{\pi^2(x)} \opS\tilde{p}_t(x)  \right]\\
        =&\partial_t \chi^2( p_t, \pi )|_{p_t =\tilde{p}_t} + \delta \partial_t \chi^2(q_t, \pi )|_{q_t =\tilde{p}_t}.
    \end{split}
    \end{equation*}
\end{small}
\hspace{-5pt} We then evaluate the above equation at the stationary distribution of the inaccurate dynamics $\tilde{\pi}$. Since $\partial_t \chi^2( \tilde{p}_t, \pi )|_{\tilde{p}=\tilde{\pi}} = 0$, we have
    \begin{equation*}
    \begin{split}
     \lambda \chi^2( \tilde{\pi}, \pi ) = \delta\left| (\partial_t \chi^2(q_t, \pi )|_{q_t =\tilde{\pi}}) \right|  <  \delta c .
    \end{split}
    \end{equation*}
\end{proof}
This theorem can also be used to measure the error in SGLD, and justifies the use of small finite step sizes in SGLD. We should note that the mixing rate bound $\lambda$ at $\tilde{\pi}$ exists for SGLD and can be obtained using spectral analysis~\cite{Levin:MCMix}, but the corresponding bounds for SGHMC are unclear due to the irreversibility of the process. We leave this for future work.

Our proof relies on a contraction bound relating the error in the transition distribution to the error in the final stationary distribution. Although our argument is based on a continuous-time Markov process, we should note that a similar guarantee can also be proven in terms of a discrete-time Markov transition kernel. We refer the reader to  \cite{Korattikara:2013austerity} and \cite{Bardenet:2014subset} for further details.

\section{Setting SGHMC Parameters}
As we discussed in Sec.~\ref{sec:sghmc-practice}, we can connect SGHMC with SGD with momentum by rewriting the dynamics as (see Eq.\eqref{eq:sghmc-normal})
\begin{equation*}
\left\{
\begin{array}{ll}
    \Delta \theta =&\hspace{-8pt} v\\
    \Delta v =&\hspace{-8pt} - \eta \nabla \psU (x) - \alpha v + \mathcal{N}( 0, 2  (\alpha-\hat{\beta})\eta ).\\
\end{array}
\right.
\end{equation*}
In analogy to SGD with momentum, we call $\eta$ the learning rate and $1-\alpha$ the momentum term. This equivalent update rule is cleaner and we recommend parameterizing SGHMC in this form.

The $\hat{\beta}$ term corresponds to the estimation of noise that comes from the gradient. One simple choice is to ignore the gradient noise by setting $\hat{\beta} = 0$ and relying on small $\eps$. We can also set $\hat{\beta} = \eta\hat{V}/2$, where $\hat{\mI}$ is estimated using empirical Fisher information as in~\cite{Ahn:2012:SGFS}.

There are then three parameters: the learning rate $\eta$, momentum decay $\alpha$, and minibatch size $|\tilde{\sD}|$. Define $\beta =\eps M^{-1}B =  \frac{1}{2} \eta \mI(\theta)$ to be the exact term induced by introduction of the stochastic gradient.  Then, we have
\begin{equation}\label{eq:param-set}
    \beta =  O\left( \eta \frac{|\sD|}{|\tilde{\sD}|} \mathcal{I} \right),
\end{equation}
where $\mathcal{I}$ is fisher information matrix of the gradient, $|\sD|$ is size of training data, $|\tilde{\sD}|$ is size of minibatch, and $\eta$ is our learning rate. We want to keep $\beta$ small so that the resulting dynamics are governed by the user-controlled term and the sampling algorithm has a stationary distribution close to the target distribution. From Eq.~\eqref{eq:param-set}, we see that there is no free lunch here: as the training size gets bigger, we can either set a small learning rate $\eta=O(\frac{1}{|\sD|})$ or use a bigger minibatch size $|\tilde{\sD}|$. In practice, choosing $\eta=O(\frac{1}{|\sD|})$ gives better numerical stability, since we also need to multiply $\eta$ by $\nabla \tilde{U}$, the mean of the stochastic gradient.  Large $\eta$ can cause divergence, especially when we are not close to the mode of distribution. We note that the same discussion holds for SGLD~\cite{Welling:2011SGLD}.

In practice, we find that using a minibatch size of hundreds (e.g $|\tilde{\sD}| = 500$) and fixing $\alpha$ to a small number (e.g. $0.01$ or $0.1$) works well. The learning rate can be set as $\eta = \gamma / |\sD| $, where $\gamma$ is the ``per-batch learning rate'', usually set to $0.1$ or $0.01$. This method of setting parameters is also commonly used for SGD with momentum~\cite{SutskeverMartensDahlHinton_icml2013}.

\section{Experimental Setup}
\subsection{Bayesian Neural Network}
The Bayesian neural network model used in Sec.~\ref{sec:BNN} can be described by the following equation:
\begin{equation}
   P( y = i | x ) \propto \exp\left(  A_i^\trs \sigma( B^\trs x + b ) + a_i \right).
\end{equation}
Here, $y\in\{1,2,\cdots,10\}$ is the output label of a digit. $A \in \rR^{10\times 100}$ contains the weight for output layers and we use $A_i$ to indicate $i$-th column of $A$. $B\in \rR^{d \times 100}$ contains the weight for the first layer. We also introduce $a\in \rR^{10}$ and $b \in \rR^{100}$ as bias terms in the model. In the MNIST dataset, the input dimension $d=784$.
We place a Gaussian prior on the model parameters
$$
    P( A ) \propto \exp( -\lambda_A \|A\|^2), P( B ) \propto \exp( -\lambda_B \|B\|^2)
$$
$$
    P( a ) \propto \exp( -\lambda_a \|a\|^2), P( b ) \propto \exp( -\lambda_b \|a\|^2).
$$
We further place gamma priors on each of the precision terms $\lambda$:
$$
    \lambda_A, \lambda_B,\lambda_a,\lambda_b \stackrel{i.i.d.}{\sim} \Gamma( \alpha, \beta ) .
$$
We simply set $\alpha$ and $\beta$ to 1 since the results are usually insensitive to these parameters. We generate samples from the posterior distribution
\begin{equation}
    P( \Theta | \mathcal{D} ) \propto \prod_{y,x\in\mathcal{D}} P( y | x, \Theta ) P(\Theta) ,
\end{equation}
where parameter set $\Theta=\{A,B,a,b,\lambda_A,\lambda_B,\lambda_a,\lambda_b\}$. The sampling procedure is carried out by alternating the following steps:
\begin{itemize}
    \item Sample weights from $P(A,B,a,b|\lambda_A,\lambda_B,\lambda_a,\lambda_b, \mathcal{D})$ using SGHMC or SGLD with minibatch of 500 instances. Sample for $100$ steps before updating hyper-parameters.
    \item Sample $\lambda$ from $P(\lambda_A,\lambda_B,\lambda_a,\lambda_b|A,B,a,b )$ using a Gibbs step. Note that the posterior for $\lambda$ is a gamma distribution by conditional conjugacy.
\end{itemize}
We used the validation set to select parameters for the various methods we compare.  Specifically, for SGD and SGLD, we tried step-sizes $\eps \in \{0.1, 0.2, 0.4, 0.8\} \times 10^{-4}$, and the best settings were found to be $\eps=0.1\times 10^{-4}$ for SGD and $\eps=0.2\times 10^{-4}$ for SGLD. We then further tested $\eps=0.16\times 10^{-4}$ and $\eps=0.06\times 10^{-4}$ for SGD, and found $\eps=0.16\times 10^{-4}$ gave the best result, thus we used this setting for SGD. For SGD with momentum and SGHMC, we fixed $\alpha=0.01$ and $\hat{\beta} = 0$, and tried $\eta \in\{ 0.1, 0.2, 0.4, 0.8 \}\times 10^{-5}$. The best settings were $\eta=0.4\times 10^{-5}$ for SGD with momentum, and $\eta=0.2\times 10^{-5}$ for SGHMC. For the optimization-based methods, we use tried regularizer $\lambda\in\{0, 0.1, 1, 10, 100\}$, and $\lambda=1$ was found to give the best performance.

\subsection{Online Bayesian Probabilistic Matrix Factorization}
The Bayesian probabilistic matrix factorization~(BPMF) model used in Sec.~\ref{sec:BPMF} can be described as:
\begin{small}
\begin{equation}
\begin{split}
    \lambda_U, \lambda_V, \lambda_a, \lambda_b &\stackrel{i.i.d.}{\sim} \mbox{Gamma}(1,1) \\
    U_{ki} \sim &\mathcal{N}( 0, \lambda_U^{-1} ), V_{kj} \sim \mathcal{N}( 0, \lambda_V^{-1} ), \\
    a_i\sim&\mathcal{N}(0,\lambda_a^{-1} ), b_i\sim\mathcal{N}(0,\lambda_b^{-1} )\\
    Y_{ij} | U, V \sim& \mathcal{N}( U_i^\trs V_j + a_i+ b_j, \tau^{-1} ).
\end{split}
\end{equation}
\end{small}
\hspace{-5pt} The $U_i\in \rR^{d}$ and $V_j\in \rR^{d}$ are latent vectors for user $i$ and movie $j$, while $a_i$ and $b_j$ are bias terms.
We use a slightly simplified model than the BPMF model considered in~\cite{SalMnih2008BPMF}, where we only place priors on precision variables $\lambda = \{\lambda_U, \lambda_V, \lambda_a, \lambda_b\}$. However, the model still benefits from Bayesian inference by integrating over the uncertainty in the crucial regularization parameter $\lambda$.
We generate samples from the posterior distribution
\begin{equation}
    P( \Theta | Y ) \propto P( Y | \Theta ) P(\Theta),
\end{equation}
with the parameter set $\Theta=\{U,V,a,b,\lambda_U,\lambda_V,\lambda_a,\lambda_b\}$. The sampling procedure is carried out by alternating the followings
\begin{itemize}
    \item Sample weights from $P(U,V,a,b|\lambda_U,\lambda_V,\lambda_a,\lambda_b, Y)$ using SGHMC or SGLD with a minibatch size of 4,000 ratings. Sample for $2,000$ steps before updating the hyper-parameters.
    \item Sample $\lambda$ from $P(\lambda_U,\lambda_V,\lambda_a,\lambda_b|U,V,a,b )$ using a Gibbs step.
\end{itemize}

The training parameters for this experiment were directly selected using cross-validation. Specifically, for SGD and $\mbox{SGLD}$, we tried step-sizes $\eps \in \{0.1, 0.2, 0.4, 0.8, 1.6 \}\times 10^{-5}$, and the best settings were found to be $\eps=0.4\times 10^{-5}$ for SGD and $\eps=0.8\times 10^{-5}$ for SGLD. For SGD with momentum and SGHMC, we fixed $\alpha=0.05$ and $\hat{\beta} = 0$, and tried $\eta \in\{ 0.1, 0.2, 0.4, 0.8\}\times 10^{-6}$. The best settings were $\eta=0.4\times 10^{-6}$ for SGD with momentum, and $\eta=0.4\times 10^{-6}$ for SGHMC.
\end{document}